\documentclass[manuscript]{aastex63}

\usepackage{url}
\usepackage{tabularx}
\usepackage{amsmath}
\usepackage{rotating}
\usepackage{multirow}
\usepackage{xcolor}
\newcolumntype{M}{>{\centering\arraybackslash}m{2.50cm}}
\newcolumntype{L}{>{\centering\arraybackslash}m{2.80cm}}

\shorttitle{Shockwaves in 3D RMHD Simulations}
\shortauthors{Sadykov et al.}

\begin{document}
	
\title{Connecting Atmospheric Properties and Synthetic Emission of Shock Waves Using 3D RMHD Simulations of Quiet Sun}
\email{sadykov@baeri.org}

\author{Viacheslav M. Sadykov}
\affiliation{NASA Ames Research Center, Moffett Field, CA 94035, USA}
\affiliation{Bay Area Environmental Research Institute, Moffett Field, CA 94035, USA}

\author{Irina N. Kitiashvili}
\affiliation{NASA Ames Research Center, Moffett Field, CA 94035, USA}
\affiliation{Bay Area Environmental Research Institute, Moffett Field, CA 94035, USA}

\author{Alexander G. Kosovichev}
\affiliation{Center for Computational Heliophysics, New Jersey Institute of Technology, Newark, NJ 07102, USA}
\affiliation{Department of Physics, New Jersey Institute of Technology, Newark, NJ 07102, USA}
\affiliation{NASA Ames Research Center, Moffett Field, CA 94035, USA}

\author{Alan A. Wray}
\affiliation{NASA Ames Research Center, Moffett Field, CA 94035, USA}

\begin{abstract}
	We analyze the evolution of shock waves in high-resolution 3D radiative MHD simulations of the quiet Sun and their synthetic emission characteristics. The simulations model the dynamics of a 12.8$\times{}$12.8$\times{}$15.2\,Mm quiet-Sun region (including a 5.2\,Mm layer of the upper convection zone and a 10\,Mm atmosphere from the photosphere to corona) with an initially uniform vertical magnetic field of 10\,G, naturally driven by convective flows. We synthesize the Mg\,II and C\,II spectral lines observed by the IRIS satellite and EUV emission observed by the SDO/AIA telescope. Synthetic observations are obtained using the RH1.5D radiative transfer code and temperature response functions at both the numerical and instrumental resolutions. We found that the Doppler velocity jumps of the C II 1334.5\,{\AA} IRIS line and a relative enhancement of the emission in the 335\,{\AA} SDO/AIA channel are the best proxies for the enthalpy deposited by shock waves into the corona (with Kendall's $\tau$ correlation coefficients of 0.59 and 0.38, respectively). The synthetic emission of the lines and extreme ultraviolet passbands are correlated with each other during the shock wave propagation. All studied shocks are mostly hydrodynamic (i.e., the magnetic energy carried by horizontal fields is $\leq{}$2.6\% of the enthalpy for all events) and have Mach numbers $>$ 1.0-1.2 in the low corona. The study reveals the possibility of diagnosing energy transport by shock waves into the solar corona, as well as their other properties, by using IRIS and SDO/AIA sensing observations.
\end{abstract}


\section{Introduction}
\label{Section:introduction}

	The solar atmosphere hosts a variety of plasma heating phenomena, including dissipation of electric currents, magnetic reconnection, and propagation of magnetohydrodynamic (MHD) shock waves. These mechanisms are typically considered as candidates to solve the coronal heating problem. However, because there are no direct in-situ plasma measurements in the solar atmosphere, the only insight into these plasma phenomena comes from remote sensing. It is therefore important to establish relationships between the properties of physical processes on the Sun and the corresponding plasma emission observed by space and ground-based telescopes. Realistic numerical simulations and synthesis of the plasma emission and spectra represent a reliable bridge for revealing such connections.
	
	The efficiency of coupling realistic simulations and synthetic modeling for diagnostic purposes was previously demonstrated in many studies. For example, the 3D radiative MHD simulations using the Bifrost code \citep{Gudiksen11a,Carlsson16a} demonstrated the diagnostic potential of the Mg\,II \citep{Leenaarts13a}, C\,II \citep{Rathore15a}, and O\,I \citep{Lin15a} lines observed by the Interface Region Imaging Spectrograph \citep[IRIS,][]{DePontieu14a}. In these studies the line synthesis was performed using the Multi3D \citep{Leenaarts09a} and RH \citep{Rybicki91a,Rybicki92a,Uitenbroek01a,Pereira15a} radiative transfer codes. \citet{Martinez-Sykora11a} investigated the formation and properties of the extreme ultraviolet (EUV) emission observed by the Atmospheric Imaging Assembly onboard the Solar Dynamics Observatory \citep[SDO/AIA,][]{Lemen12a}. \citet{Kitiashvili15a} investigated the properties of the Fe\,I\,6173\,{\AA} line observables delivered by the Helioseismic and Magnetic Imager onboard the Solar Dynamics Observatory \citep[SDO/HMI,][]{Scherrer12a} using realistic modeling of the solar convection zone and photosphere using the StellarBox code \citep{Wray15a,Wray18a} and the radiation transfer code SPINOR/STOPRO \citep{Solanki12a}. \citet{Bjorgen19a} studied the formation of various lines (H$\alpha$, Mg\,II\,k, Ca\,II\,K, and Ca\,II\,8542\,{\AA}) using MURAM and Multi3D codes for realistic 3D MHD simulations of solar flares \citep{Cheung19a}. Thus the pairing of realistic MHD modeling and synthetic spectral calculations is a powerful approach to study phenomena in the solar atmosphere, including MHD waves and shocks.
	
    From an observational perspective, high-resolution, high-cadence spectroscopic observations provide a promising opportunity for wave and shock diagnostics. Using high-cadence ``sit-and-stare'' observations by the IRIS satellite, \citet{Tian14a} analyzed the behavior of shocks in sunspot atmospheres. In particular, the authors found that the Mg\,II, C\,II, and Si\,IV spectral lines experience periodic intensity peaks, Doppler shift changes, and line-width enhancements associated with the propagation of sunspot oscillations in the higher atmosphere and the formation of shocks in the solar corona. Similar observational patterns of various spectral lines were also reported in other work \citep{Centeno06a,Chae15a,Kanoh16a,Skogsrud16a,Anan19a,Houston20}. There were attempts to model such patterns. For example, the formation of the Ca\,II\,H$_{2v}$ bright grains by acoustic shocks was modeled by \citet{Carlsson97a} by solving the radiative hydrodynamic equations with a detailed atomic excitation/ionization model. The authors reached good agreement between observations and models and concluded that the bright grains are produced primarily by waves traveling from the photosphere, the frequencies of which are slightly above the acoustic cutoff frequency. \citet{Ruan18a} developed an analytical framework for shockwave analysis based on the single-fluid Rankine-Hugoniot relations, and derived upstream and downstream velocities from the observed Si\,IV line parameters for the data set used by \citet{Tian14a}. The authors also found good agreement between the observations and models.
	
    Recent realistic simulations using the StellarBox code \citep{Wray15a,Wray18a} demonstrated the formation of shocks in the quiet-Sun atmosphere and corona \citep{Kitiashvili20a}. In this work, we investigate how various properties of the shocks are connected to properties of the synthesized emission. The simulation setup and computational procedure are discussed in Section~\ref{Section:stellarbox}. Physical properties of the shocks and corresponding emission and atmospheric properties are discussed in Section~\ref{Section:shocks}. The relationship between the atmospheric properties of shock waves and the corresponding emission is illustrated in Section~\ref{Section:results}, followed by a discussion of results in Section~\ref{Section:discussion} and main conclusions in Section~\ref{Section:conclusions}.

\section{Synthetic Spectra and EUV Emission for 3D RMHD Simulations}
\label{Section:stellarbox}

	\subsection{Description of RMHD Simulations}
	
		The modeling is performed using the three-dimensional radiative magnetohydrodynamic (RMHD) code ``StellarBox'' \citep{Wray15a,Wray18a}. The code solves the compressible MHD equations on a three-dimensional Cartesian grid and includes a fully-coupled radiation solver in the local thermodynamic equilibrium (LTE) approximation. The code also incorporates the Smagorinsky turbulence model for subgrid turbulent transport. Originally developed for the modeling of magnetoconvection, photospheric, and chromospheric phenomena \citep{Jacoutot08a,Kitiashvili12a,Kitiashvili15a,Kitiashvili15b,Kitiashvili19a}, the code's capabilities have been extended to model the solar corona.
		
		In the current work, the computational domain has a size of 12.8$\times{}$12.8$\times{}$15.2\,Mm, which includes 10\,Mm of the solar atmosphere. The spatial resolution is 25\,km in horizontal directions. The vertical grid-spacing varies from 13\,km at the photosphere to 76\,km in the solar corona, with a total of 512 vertical grid cells. The standard solar interior model by \citet{Christensen-Dalsgaard96a} and static quiet-Sun VAL model by \citet{Vernazza81a} are utilized for the initial setup of the interior and atmospheric stratifications. The initial vertical uniform magnetic field is 10\,G across the computational domain. The mean magnetic flux is maintained by the boundary conditions. The coronal temperature was maintained at about 1\,MK by artificial coronal heating at heights above 2\,Mm.
		
\begin{figure}[t!]
	\centering
	\includegraphics[width=0.95\linewidth]{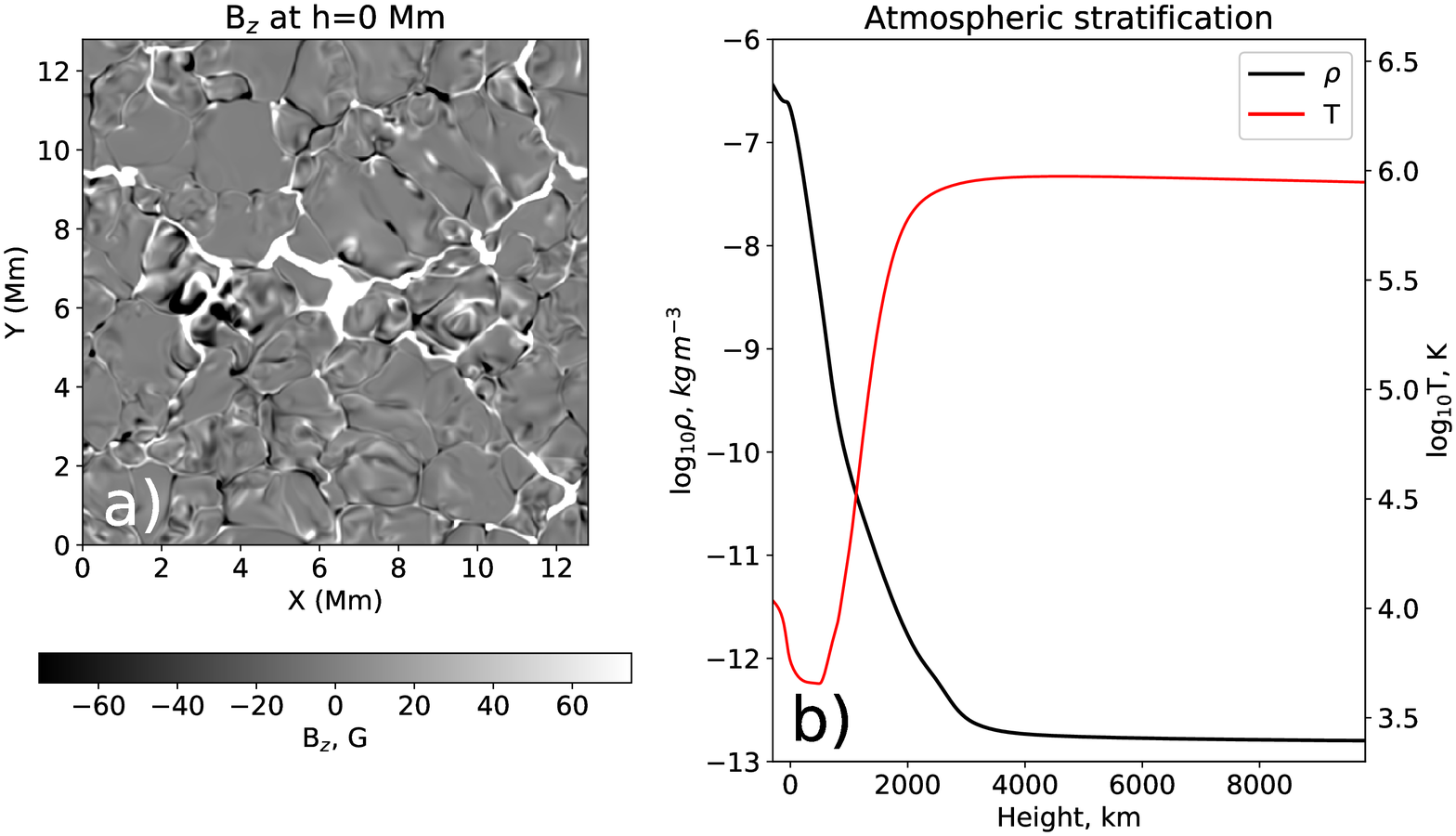} \\
	\caption{Illustration of (a) a vertical magnetic field map at z = 0\,Mm height, and (b) a stratification of the horizontal mean density and temperature for the initial time moment of the considered simulation series.}
	\label{figure0}
\end{figure}

		After relaxation of the MHD flow was reached, the artificial coronal heating was switched off. A distribution of the photospheric vertical magnetic field and a stratification of the horizontally-averaged density and temperature for the initial time moment of the utilized simulation series is illustrated in Figure~\ref{figure0}. The simulations resulted in a hotter chromosphere and denser corona in comparison with the initial VAL model. One can also notice a strong positive magnetic field patch in the middle of the computational domain, with the actual magnetic field values in the patch of about 1\,kG, formed spontaneously from the initial uniform 10\,G vertical magnetic field. Such a strong photospheric magnetic field concentration determines the global structuring of the simulated corona. This is evident in Figures~\ref{figure1}\textit{a}~and~\ref{figure1}\textit{b}. One can see a funnel-like structure in the temperature and density distributions at 4\,Mm height persistent throughout the duration of the considered simulation series. More details about this particular simulation run can be found in \citet{Kitiashvili20a}.

	\subsection{Synthesis of Spectral Lines Observed by IRIS}
	
		The Mg\,II\,k\,\&\,h and C\,II\,1334\,\&\,1335\,{\AA} spectral lines observed by IRIS are thought to originate in the upper chromosphere (the characteristic formation temperature is 8-10$\times{}$10$^{3}$\,K) and in the lower transition region \citep[1.4-5.0$\times{}$10$^{4}$\,K,][]{Rathore15a}. To synthesize these lines, we utilize the RH1.5D code \citep{Pereira15a}~--- the massively-parallel version of the RH code \citep{Rybicki91a,Rybicki92a,Uitenbroek01a}. This code solves the atomic population equation under the statistical equilibrium assumption and provides column-by-column computations assuming the plane-parallel approximation for the atmosphere within each column. The code input is, in essence, the 3D distribution of temperature, density, vertical velocity, height scale, and electron number density computed in local thermodynamic equilibrium (LTE) conditions using the OP project data \citep{Seaton94a}. For the analysis we use a 6-min time series of the simulation with 2\,s cadence. The number density of electrons is not recalculated and is forced to be in LTE. This assumption does not lead to any qualitative changes of the Mg\,II and C\,II line profiles which are calculated in NTLE approximation.
		
        We perform calculations in two steps. First, we solve the radiative transfer problem for the H and Mg atoms, with all other essential species (He, O, C, N, Fe, Si, S, Al, Ca, Na, Ni) kept in LTE. Second, we solve the problem solely for the C species while keeping the Hydrogen population from the previous solution. The C\,II line profiles are solved under the complete frequency redistribution approximation \citep[the applicability of this assumption is discussed in][]{Rathore15b}. The Mg\,II\,h\,\&\,k line profiles are computed assuming the partial frequency redistribution (PDR) approximation. As a result of the synthetic calculations, for each time moment we obtain the Mg\,II and C\,II spectra with the high (25\,km) spatial resolution of the simulation and with a spectral resolution exceeding the instrumental (IRIS) resolution by about 10 times.
		
		Examples of the computed line profiles are presented in Figures~\ref{figure1}\textit{g}~and~\ref{figure1}\textit{h} by dashed lines. Point 1 in this figure indicates a region of enhanced temperature and density outside a self-organized magnetic structure; Point 2 samples plasma conditions inside the structure. The small-scale magnetic structure is formed spontaneously from the initially-uniform magnetic field. The horizontal coordinates are chosen to place this structure in the middle. One can see that both the C\,II and Mg\,II lines demonstrate a central-reversal signature which indicates that they are optically thick.
	
	\subsection{Synthesis of EUV Emission Observed by SDO/AIA}
	
		The high-temperature plasma of the solar corona generates strong emission in the extreme ultraviolet (EUV) range that is observed by SDO/AIA. We model the emission of the 7 EUV channels of SDO/AIA in the optically-thin assumption utilizing the temperature response functions documented in the IDL Solar SoftWare package (SSW). For a given temperature, the temperature response function gives the contribution of a unity-emission-measure plasma element to the overall EUV emission. SDO/AIA emission is computed for 94\,$\AA$, 131\,$\AA$, 171\,$\AA$, 193\,$\AA$, 211\,$\AA$, 304\,$\AA$, and 335\,$\AA$ by integration over the vertical direction of emission measure in each plasma element (Fig.~\ref{figure1}\textit{c,e}).
	
	\subsection{Reduction of Resolution from Computational to Instrumental}
	
		Although we have the advantage of analyzing shock waves at high spatial and spectral resolutions, it is important to look at their properties as ``observed'' by real instruments. For such purposes, we degrade the resolution of the computed emission to the corresponding instrumental resolution \citep{Lemen12a,DePontieu14a}. For both the spatial and spectral resolution reductions, we assume that the Point Spread Function (PSF) is Gaussian and that ``resolution'' corresponds to full width at half maximum. In particular:
		\begin{itemize}
			\item The spectral resolution of the Mg\,II line profiles is degraded from $\sim$3\,m{\AA} to 53\,m{\AA} with a 25.6\,m{\AA} wavelength spacing. The spatial resolution is degraded from $\sim$0.035'' to 0.40'' with a 0.167'' pixel size;
			\item For the C\,II line profiles, the spectral resolution is reduced from $\sim$3\,m{\AA} to 26\,m{\AA} with a 12.8\,m{\AA} wavelength spacing. The spatial resolution is reduced from $\sim$0.035'' to 0.33'' with a 0.167'' pixel size;
			\item The spatial resolution of the SDO/AIA emission is degraded from $\sim$0.035'' to $\sim$1.50'' (the exact number depends on the channel) with a 0.60'' pixel size;
		\end{itemize}
		
\begin{figure}[t!]
	\centering
	\includegraphics[width=0.95\linewidth]{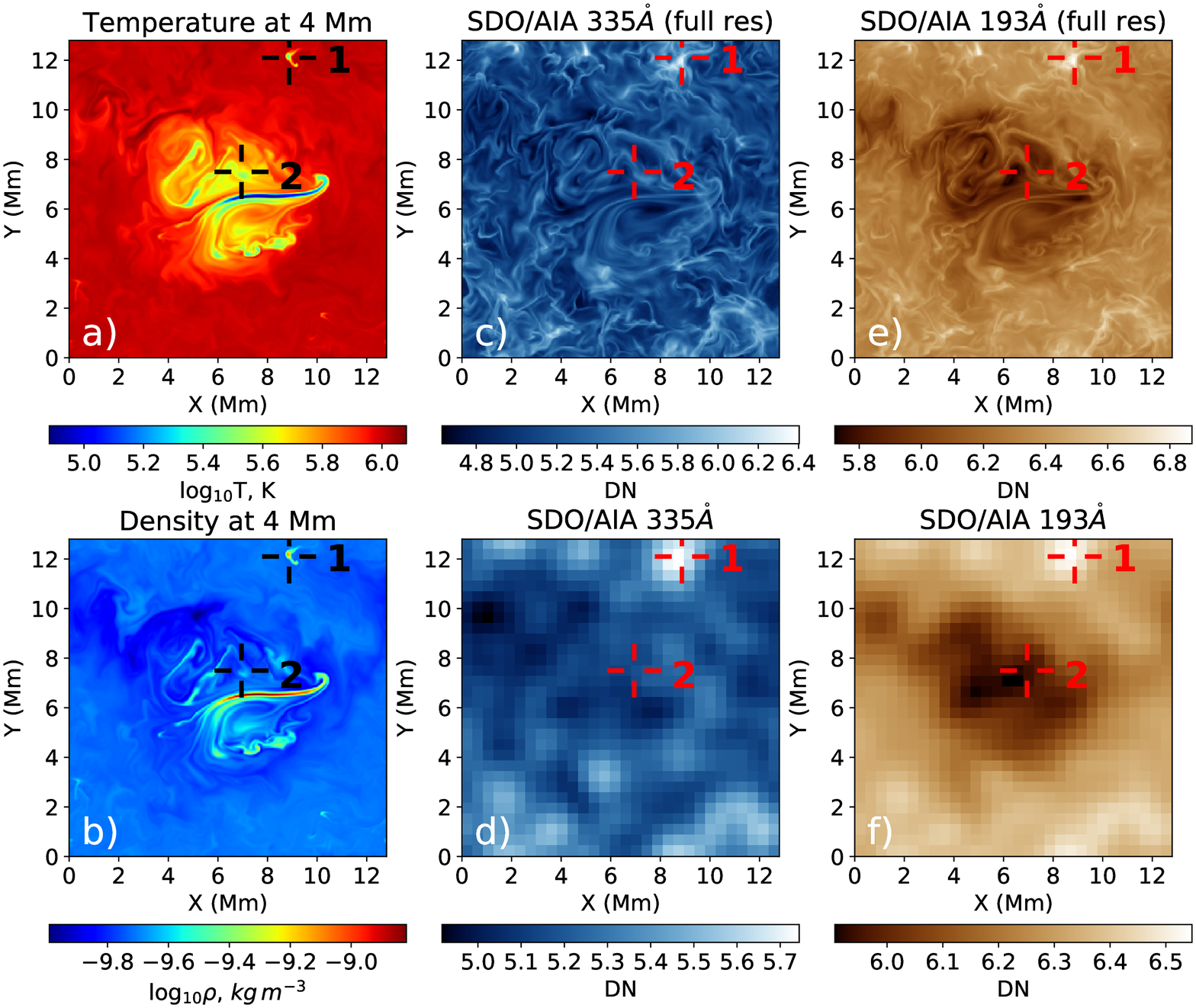} \\
	\vskip -15pt
	\includegraphics[width=0.95\linewidth]{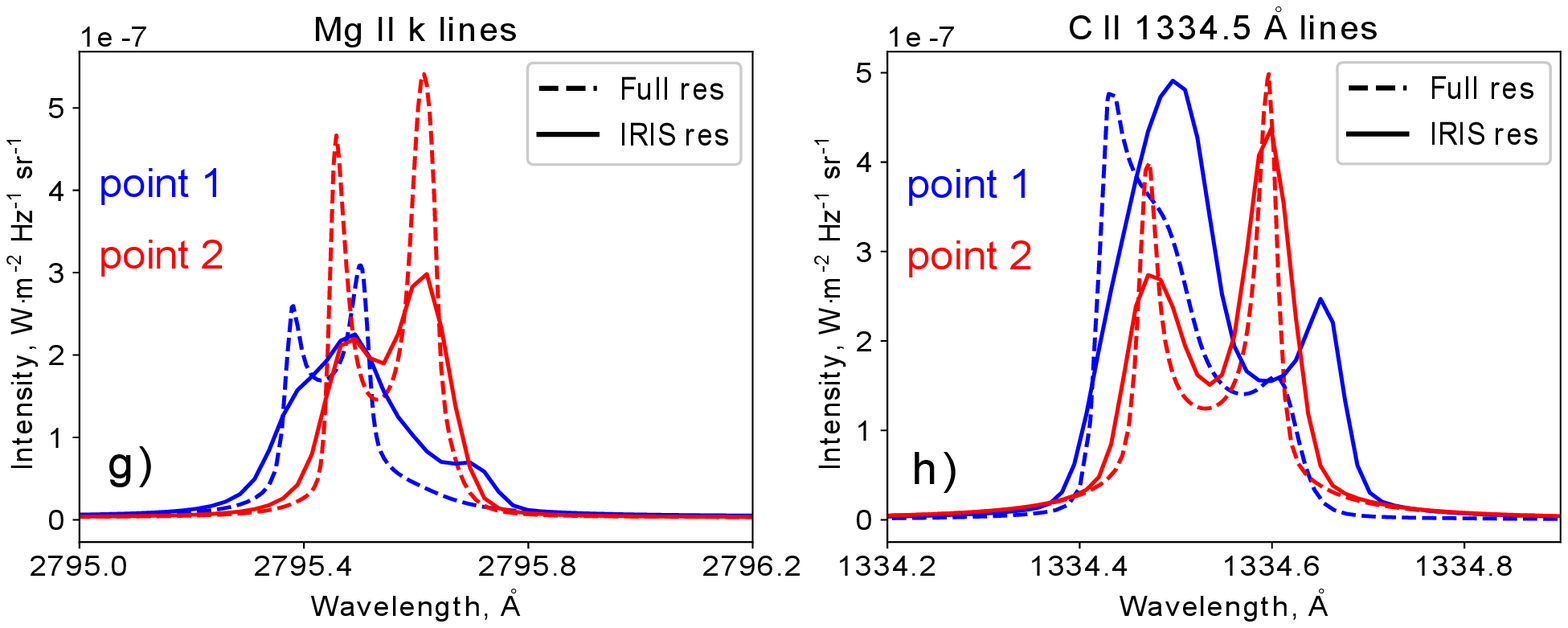}
	\caption{Illustration of physical properties and synthetic emission for the initial time moment of the considered simulation series: (a) temperature and (b) density distributions at 4\,Mm height; synthetic SDO/AIA 335\,{\AA} emission at (c) computational and (d) instrumental resolutions; synthetic SDO/AIA 193\,{\AA} emission at (e) computational and (f) instrumental resolutions; (g) Mg\,II\,k and (h) C\,II\,1334.5\,{\AA} line profiles derived in points 1 (blue) and 2 (red) at computational (dashed) and instrumental (solid) resolutions. Points 1 and 2 are marked by targets in panels a-f.}
	\label{figure1}
\end{figure}

		Examples of emission for the instrumental resolution at the beginning of the analyzed simulation series are presented in Figures~\ref{figure1}\textit{d}~and~\ref{figure1}\textit{f} (for SDO/AIA images) and in Figures~\ref{figure1}\textit{g}~and~\ref{figure1}\textit{h} by solid line profiles (for IRIS spectral lines). A significant difference between the high-resolution and low-resolution Mg\,II and C\,II line profiles is caused by the reduction of the spatial resolution from computational to instrumental.

\section{Manifestation of Shock Waves in Simulations}
\label{Section:shocks}

	In this Section, we describe the detection procedure for shocks and their emission and physical properties.

	\subsection{Manual Detection of Shock Waves}
	
\begin{sidewaysfigure}[t!]
	\centering
	\includegraphics[width=1.0\linewidth]{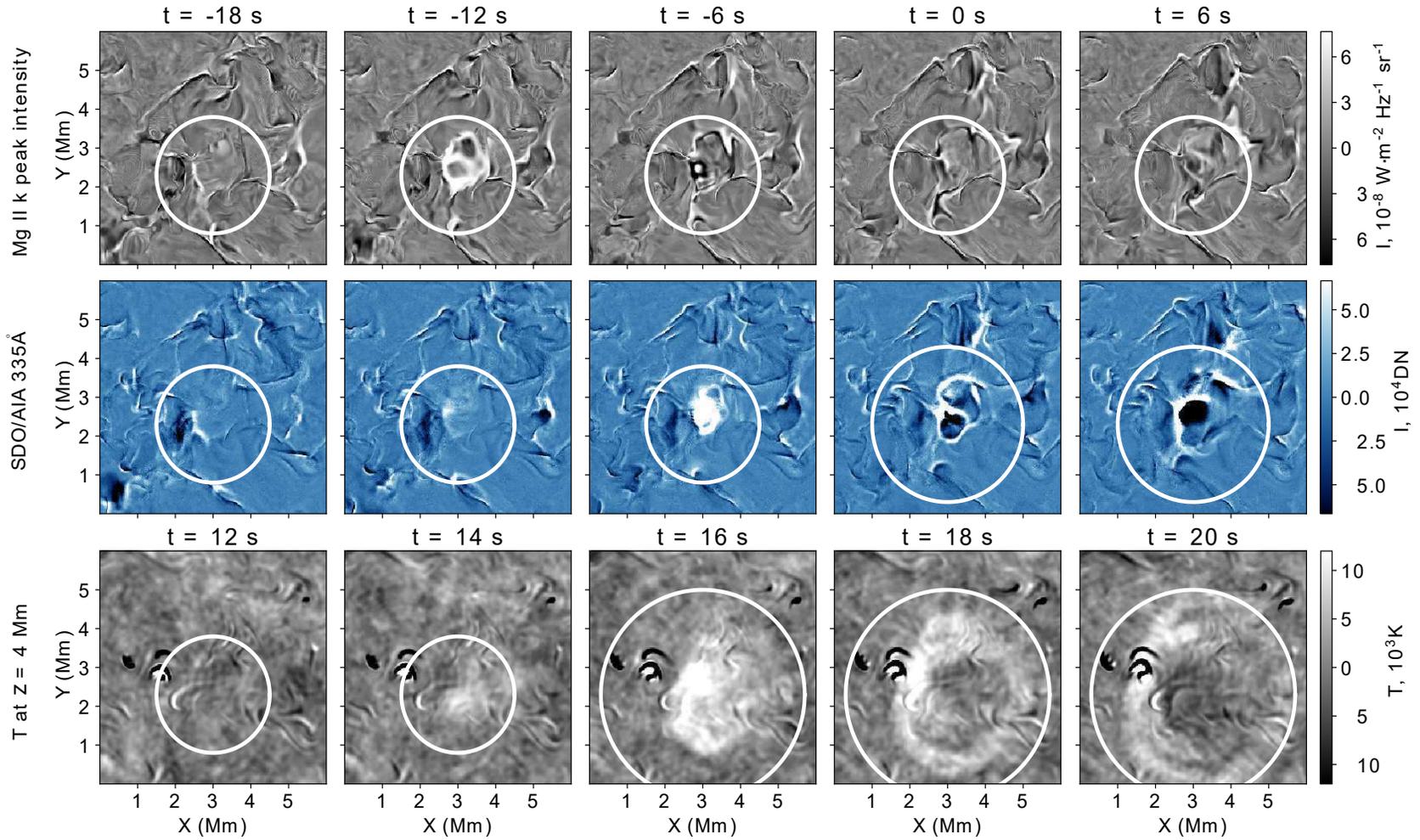}
	\vskip -20pt
	\caption{Observational signatures of the shock event shown in Figure~\ref{figure4} revealed in running difference images of synthetic Mg\,II\,k line peak intensity (top row), SDO/AIA 335\,{\AA} emission (middle row), and distribution of temperature at 4\,Mm (bottom row). The time moments are defined with respect to the shock reference time. The region where the shock appears is marked by a circle.}
	\label{figure2}
\end{sidewaysfigure}

		The presence of shocks in the simulation series is evident both in the physical properties of the atmosphere and in the synthesized emission. The perturbations ultimately causing shocks originate in the lower atmosphere and propagate upward, which is evident from the analysis of physical parameters. When reaching the transition region, the perturbations accelerate and, finally, penetrate into the solar corona. As will be demonstrated later, the speed of the waves in the corona is slightly larger than the local speed of sound and an order of magnitude larger than velocities of the local hydrodynamic flows. We found that the shocks can be identified as localized expanding enhancements in the synthetic SDO/AIA 335\,{\AA} images, accompanied later on by temperature perturbations at 4\,Mm, as illustrated in Figure~\ref{figure2} for the selected event.
		
		Figure~\ref{figure2} illustrates the evolution of a selected shock wave propagation in Mg\,II peak intensity, 335\,$\AA$ emission, and temperature at 4\,Mm height. Using the periodic lateral boundary conditions, we rearranged the computational domain for better visibility of the shock. The shocks typically manifest themselves as enhancements in all SDO/AIA channels. However, the enhancements in the 94\,{\AA} and 171\,{\AA} channels are less prominent and not found for every event. Initial enhancements of shock waves in 335\,$\AA$ have been identified manually and referred to as ``shock centers'' hereafter. Perturbations related to the selected shock are visible in the middle row of the Figure~\ref{figure2} where the running difference images are shown for the SDO/AIA 335\,{\AA} synthetic emission. Perturbations related to shock wave propagation are also visible at the peak intensities of the computed Mg\,II lines, as illustrated in Figure~\ref{figure2} (top row) for Mg\,II\,k line. The bottom row of Figure~\ref{figure2} also illustrates the corresponding running difference of temperature at 4\,Mm height. The circular-like perturbation becomes visible approximately 18 seconds after perturbations in UV lines and EUV emission, as evident in the last two panels of the row.
		
        Figures~\ref{figure4}\textit{a-j} show the vertical velocity running differences along two orthogonal vertical planes centered at the shockwave event. The time moments in the panels are defined relative to the strongest emission in 335\,$\AA$ observed during the shock propagation. Figure~\ref{figure4}\textit{k} presents the vertical time-distance (TD) diagram for vertical velocity differences of the selected event. The diagram is obtained by integration of the horizontal slices within $\pm$250\,km from the shockwave center. One can notice the prominent white ridge there. By estimating its slope at heights of 2-5\,Mm, one can estimate the speed of propagation of the shockwave front in the vertical direction in the corona. Because in the corona the density decreases with height very slowly in comparison with the chromosphere and transition region, we can assume that the propagation of the shock occurs as in an unstratified medium and estimate the vertical speed of its front, $v_{z}^{f}$, from the TD-diagram shown. One can see that the ratio of $v_{z}^{f}$ to the average speed of sound at 2-5\,Mm, $c_{s}$, ranges from 1.0 to 1.2 for the analyzed events (Fig.~\ref{figure4}\textit{l}). Because the shock front may not be traveling exactly in the vertical direction, as evident in Figure~\ref{figure4}\textit{c,h}, the derived $v_{z}^{f}$/$c_{s}$ values represent lower limits of the Mach numbers of the shocks. Also, the hydrodynamic velocities in the vicinity of the propagating shock front are typically an order of magnitude lower than the vertical speed of the shock front $v_{z}^{f}$.
        
        To summarize, a total of 18 shocks were detected across the computational domain in the 6-min simulation series. For convenience, we define the reference time of each shock as the time of the maximum enhancement in 335\,{\AA} EUV emission in the shock center, though the peak enhancements can occur at different times for different lines and EUV emission channels.
	
	\subsection{Typical Behavior of Synthetic Observables and Physical Parameters of Shocks}
		
		Figure~\ref{figure3} shows the temporal behavior of the synthesized emission and physical parameters in the center of the shock wave event illustrated in Figure~\ref{figure2}. One can see that the event is prominent in the Mg\,II and C\,II spectra (Fig.~\ref{figure3}\textit{a}~and~\ref{figure3}\textit{b}): the line emission is enhanced, and the line profiles experience a strong Doppler shift change at the time of shockwave propagation. Both the Mg\,II and C\,II line profiles are redshifted before the event and blueshifted after it. Note that the synthetic data are considered here using the instrumental resolution. Such signatures in the observed IRIS spectra were previously reported for the same lines \citep{Tian14a,Ruan18a}. One more notable feature of the spectra, visible in both considered lines but especially prominent in the C\,II\,1334.5\,{\AA} line (Fig.~\ref{figure3}\textit{b}), is a flip of the line intensity maximum from the shorter-wavelength line peak to the longer-wavelength one. We note that this feature appears for the majority of the shocks analyzed in this paper.
		
		Figure~\ref{figure3}\textit{c} illustrates the SDO/AIA intensity enhancement during shockwave propagation. Depending on the SDO/AIA channel, the emission can be enhanced up to three times with respect to the pre-shock values. One can also see that the emission in the 304\,{\AA} channel (which has a primary contribution from the lower temperature plasma compared to the other channels) is slightly delayed with respect to the 193\,{\AA} and 335\,{\AA} channels.
		
        Figure~\ref{figure3}\textit{d} illustrates the behavior of the enthalpy flux and the vertical velocity at the height corresponding to the chromosphere-corona transition region with a temperature of 5$\times{}$10$^{5}$\,K. The enthalpy flux is averaged over the SDO/AIA PSF full width at half maximum (FWHM, $\sim$1.50''). One can see that the shock wave causes an inflow of enthalpy into the corona. We also see a sharp change in the vertical velocity from downward to upward.
		
        Although Figures~\ref{figure2}-\ref{figure3} illustrate one shockwave example, we have confirmed similar behavior for the majority of the 18 selected shockwave events: all of them had the discussed signatures in the UV spectra and EUV emission, and most of them were accompanied with enthalpy inflow into the corona (except one event where there was no upward velocity detected at the 5$\times{}$10$^{5}$\,K height after the shock) and had a detectable shockwave front ridge in the vertical TD diagrams (except two events when the detection of such a ridge was ambiguous). In the next section, we quantify the shock properties and analyze relationships between their physical characteristics and emission properties.
	
	\subsection{Physical Characteristics and Synthetic Emission Parameters of Shocks}
	
		To quantify shock wave propagation, we define and derive the following characteristics at the heights corresponding to 1$\times{}$10$^{4}$\,K, 2$\times{}$10$^{4}$\,K, and 5$\times{}$10$^{5}$\,K in the upper chromosphere and transition region:
		\begin{itemize}
			\item Enthalpy and magnetic energy inflows integrated for 60\,s after the shockwave reference time (defined as the time of strongest enhancement of SDO/AIA 335\,{\AA} emission). The fluxes are first computed with the numerical resolution and then averaged over the SDO/AIA PSF FWHM;
			\item The difference between the maximum and minimum velocities (hereafter ``velocity jump at 1$\times{}$10$^{4}$\,K'', `` velocity jump at 2$\times{}$10$^{4}$\,K'', and ``velocity jump at 5$\times{}$10$^{5}$\,K''), averaged over the SDO/AIA PSF FWHM;
			\item Same parameters as above but for the computational (25\,km) spatial resolution in the vicinity of the identified shock center;
			\item The ratio of the vertical speed of the shock front at the corona, $v_{z}^{f}$, to the average speed of sound at 2-5\,Mm, $c_{s}$.
		\end{itemize}
		
		To quantify the emission properties, we derive the following characteristics computed for the instrumental resolutions starting from 30\,s before the shock reference time and ending 60\,s after the shock reference time:
		\begin{itemize}
			\item The ratio of the largest and smallest Mg\,II\,k and C\,II\,1334.5\,{\AA} peak intensities during shock propagation  (enhancement ratios);
			\item Difference between the largest positive and negative Doppler shifts for the Mg\,II\,k and C\,II\,1334.5\,{\AA} lines (Doppler shift jump) derived using the center-of-gravity approach \citep{Sadykov19a};
			\item Enhancement ratios for the SDO/AIA channels;
			\item The same parameters as above but for the computational (25\,km) spatial resolution;
		\end{itemize}
		
		In addition to these parameters, we measure the time intervals between the emission enhancements and the centers of the Doppler shift jumps (the time moments when the Doppler shifts cross zero). Physical characteristics and synthetic emission parameters of shocks are summarized in Tables~\ref{table1}~and~\ref{table2}. Relationships among the parameters are examined using a correlation analysis described in the following section.

\section{Correlation Analysis of Shock Characteristics and Synthetic Emission Properties}
\label{Section:results}

	To analyze correlations between the physical characteristics of the shock waves and the corresponding synthesized emission properties, we calculate the non-parametric Kendall's $\tau$ coefficient (Kendall's rank correlation coefficient). For datasets of the same size with two parameters, $\{x\}$ and $\{y\}$, Kendall's $\tau$ is defined as:
	
	\begin{gather}
	\label{eq:Kendallstau}
	\tau{} = \dfrac{2}{n(n-1)}\sum_{i<j}sgn(x_{i}-x_{j})sgn(y_{i}-y_{j})
	\end{gather}
	
    Here $sgn$ is the sign function, and $n$ is the number of elements in each data set. Kendall's $\tau$ ranges between -1 and 1; its value is expected to be 0 for independent data sets. To indicate whether the obtained correlation coefficients are statistically significant, we also look at the p-value for a hypothesis test whose null hypothesis is an absence of correlation ($\tau{}=0$). P-values represent the probability to incorrectly deduce the presence of a correlation based on the given data sampling. Large p-values ($>$ 0.05) indicate that no strong conclusions can be made with respect to the derived $\tau$ value.
    
    Figure~\ref{figure5} illustrates the correlations between the enthalpy deposited by the shocks into the corona (the enthalpy flux integrated over positive values within 1\,min after the shockwavereference time) and the synthesized emission parameters derived for the instrumental resolution. We display the parameters having the strongest correlations with the enthalpy deposit: the Mg\,II\,k line maximum intensity enhancement and Doppler shift jump (Fig.~\ref{figure5}\textit{a-b}), the C\,II\,1334.5\,{\AA} line maximum intensity enhancement and Doppler shift jump (Fig.~\ref{figure5}\textit{c-d}), and enhancements of the SDO/AIA 335\,{\AA} and 193\,{\AA} emissions (Fig.~\ref{figure5}\textit{e-f}). All other parameters were also examined but demonstrated weaker correlations.
	
    Correlations between the C\,II\,1334.5\,{\AA} Doppler shift jump and the enthalpy deposit are the strongest found in this study: the corresponding Kendall's $\tau$ value is $\tau{}=0.59$ and has the strong statistical significance (p-value $<$ 0.001, see Figure~\ref{figure5}\textit{d}). In the framework of this study, the C\,II\,1334.5\,{\AA} Doppler shift jump is the best proxy for the enthalpy transported by shocks into the corona. The Mg\,II\,k line parameters, together with the enhancement of the C\,II peak intensity relative to the pre-shockwave value, have lower $\tau$ values, although they do demonstrate statistically-significant correlations. It is interesting that the synthesized SDO/AIA 335\,{\AA} emission enhancement relative to its pre-shockwave value demonstrates clear correlation with the enthalpy deposit, with $\tau{}=0.38$. Properties of other SDO/AIA channels sensitive to the emission of high-temperature plasma demonstrate significantly lower correlations: for example, for the SDO/AIA 193\,{\AA} emission properties, the $\tau$ value drops to $\tau{}=0.28$, and the corresponding p-value exceeds the 0.05 threshold.
	
	Figure~\ref{figure6} illustrates other selected correlations for the shockwave parameters. In particular, Figure~\ref{figure6}\textit{a} illustrates the correlation between the enthalpy flux at heights corresponding to  1$\times{}$10$^{4}$\,K and 5$\times{}$10$^{5}$\,K as defined before the shockwave propagation. Note here that the enthalpy flux at the height of 5$\times{}$10$^{5}$\,K includes the gravitational potential energy difference with respect to the 1$\times{}$10$^{4}$\,K height, which gives just slightly higher deposited fluxes with respect to those presented in Figure~\ref{figure5}. One can see that the enthalpy flux decreases by about 33\% (the median value) in the transition region. Figure~\ref{figure6}\textit{b} demonstrates that the velocity maxima are correlated with the velocity jumps at the same height. Figure~\ref{figure6}\textit{c} demonstrates the correlations between the C\,II\,1334.5\,{\AA} Doppler jump and the velocity jump at the 5$\times{}$10$^{5}$\,K height. One can see a strong correlation even though the C\,II line typically originates at lower temperatures. Such correlations will be an essential point of discussion in Section~\ref{Section:discussion}. Figure~\ref{figure6}\textit{d} presents the correlation of the C\,II\,1334.5\,{\AA} Doppler jump and the vertical velocity jump at the 2$\times{}$10$^{4}$\,K pre-shock height. As one can see, these values are very close to each other in magnitude, although the C\,II Doppler shift jumps have slightly lower values. This indicates that the C\,II Doppler shift jump computed with the Center-of-Gravity method is a relatively good proxy for the vertical velocity jump. Figure~\ref{figure6}\textit{e} highlights the correlation between the enhancement ratios of the C\,II line and the SDO/AIA\,335\,{\AA} emission, indicating the correlated behavior of two different types of measurements. Finally, Figure~\ref{figure6}\textit{f} illustrates that the shock velocity jump obtained at the T = 10$^{4}$\,K height correlates with the time difference between the SDO/AIA 335\,{\AA} enhancement and the Mg\,II\,k Doppler shift jump. We have to mention that, although the event illustrated in Figure~\ref{figure3} demonstrates the sequential appearance of the Mg\,II, C\,II, SDO/AIA 304\,{\AA} and 335\,{\AA} peak intensities, the correlation presented in Figure~\ref{figure6}\textit{f} is the only statistically-significant correlation between the timing properties and physical properties of the shocks found in this study.
	
    There are also several other results not highlighted in the presented figures that are important to mention. First, the magnetic energy carried by horizontal magnetic fields, $\dfrac{B_{h}^{2}}{8\pi{}}$, is less than 2.6\% of the enthalpy for all considered events. As a result, the vertically-propagating shocks are predominantly hydrodynamic (the component of the magnetic field perpendicular to the shockwave front, B$_{z}$ in our case, does not experience a jump). Also, the mean value of the full magnetic energy, $\dfrac{B^{2}}{8\pi{}}$, is about 7\% of the enthalpy.
	
	In addition to the correlations for the instrumental resolution, we investigated correlations of the parameters obtained at the high (computational) resolution within the 250\,km x 250\,km region around the shock center. The parameters obtained with high-resolution demonstrate weaker correlations. For example, the median value of Kendall's $\tau$ for correlation between the enthalpy deposit and the C\,II Doppler shift jump is equal to $\tau{}=0.38$. Correlations for the high-resolution parameters of the Mg\,II line become not statistically significant anywhere around the shock: the corresponding p-values are significantly higher than the 0.05 threshold typically assumed to claim a statistically significant correlation. The same stands for correlations between the enthalpy deposit and SDO/AIA emission properties.

\section{Discussion}
\label{Section:discussion}

    The shock waves studied in the radiative MHD simulations of the quiet Sun are of complex three-dimensional nature. This is evident, for example, in Figure~\ref{figure4}\textit{a-j} where the shockwave event has a complex shape, especially in its initial phase. Our analysis demonstrates that the properties of the synthesized emission correlate with the changes in the atmospheric parameters during shockwave propagation. This is very important for the potential development of shockwave diagnostic techniques in observational data. Although the signatures of shock waves may not be clear from observations of individual lines and emission channels (partially due to generally low intensities of lines and SDO/AIA emission in quiet-Sun regions combined with a high level of noise), the key to identify shock wave events is in the correlated enhancements of the line and emission properties. The synthetic emission properties of shocks (C\,II and Mg\,II line enhancements and Doppler shift jumps, and SDO/AIA emission enhancements) correlate with each other. When a shock propagates in the atmosphere, enhancements of these parameters are correlated, as illustrated, for example, in Figure~\ref{figure6}e. Also, the timing properties may correlate with each other as seen in Figure~\ref{figure6}f.
	
	Our analysis of the synthetic observations shows a strong correlation of the C\,II\,1334.5\,{\AA} Doppler shift jump and the enthalpy deposited into the corona. Although such a correlation is intuitively reasonable, it is also easy to justify it. As mentioned above, the shockwaves are mostly hydrodynamic (the fraction of enthalpy carried by the horizontal magnetic field is $<$2.6\%). In this case, assuming local homogeneity of the medium, one can write the Rankine-Hugoniot equations for the single-fluid uniform medium as:
	\begin{gather}
	\label{eq:continuity}
	\rho{}_{1}u_{1}=\rho{}_{2}u_{2} \\
	\rho{}_{1}u_{1}^{2} + p_{1}=\rho{}_{2}u_{2}^{2} + p_{2} \\
	\rho{}_{1}u_{1}\epsilon{}_{1} + p_{1}u_{1} + \rho{}_{1}u_{1}^{3}/2 = \rho{}_{2}u_{2}\epsilon{}_{2} + p_{2}u_{2} + \rho{}_{2}u_{2}^{3}/2
	\end{gather}
	
	Here $u_{1}$, $\rho{}_{1}$, $\epsilon{}_{1}$, and $p_{1}$ correspond to the upstream velocity, density, internal energy, and pressure, in the rest frame of the shock, and $u_{2}$, $\rho{}_{2}$, $\epsilon_{2}$, and $p_{2}$ are the downstream values. The difference in the velocities before and after the shock, $v_{j}$, does not depend on the reference frame, so one can write:
	\begin{gather}
	\label{eq:downstream}
	u_{2} = u_{1} - v_{j}
	\end{gather}
		
	The relations between the upstream and downstream parameters of the shock wave are given by the Rankine-Hugoniot relations as \citep{Ruan18a}:
	\begin{gather}
	\label{eq:densityrel}
	\dfrac{\rho{}_{2}}{\rho{}_{1}} = \dfrac{(\gamma{}+1)M_{1}^{2}}{(\gamma{}-1)M_{1}^{2}+2} \\
	\label{eq:temperaturerel}
	\dfrac{T_{2}}{T_{1}} = 1 + \dfrac{2(\gamma{}-1)}{(\gamma{}+1)^{2}} \dfrac{\gamma{}M_{1}^{2}+1}{M_{1}^{2}}(M_{1}^{2}-1)
	\end{gather}
		
	Here $T_{1}$ and $T_{2}$ correspond to the temperatures of the upstream and downstream flows, $M_{1}$ is the Mach number for the upstream medium, and $\gamma{}$ is the adiabatic exponent. By combining Eq.~\ref{eq:continuity}, \ref{eq:downstream}, and \ref{eq:densityrel}, one can find the relation between the velocity jump, $v_{j}$, and the shock speed relative to the upstream medium (which is equal to $u_{1}$):
	\begin{gather}
	\label{eq:veljump}
	v_{j} = c_{s1}M_{1}\dfrac{M_{1}^{2}+1}{(\gamma{}+1)M_{1}^{2}}
	\end{gather}
		
	Here $c_{s1}$ is the speed of sound in the upstream conditions. One can check that, for any $M_{1} > 1$:
	\begin{gather}
	\dfrac{\partial{(\rho{}_{2}/\rho{}_{1})}}{\partial{M_{1}}} > 0, ~~~~~
	\dfrac{\partial{(T_{2}/T_{1})}}{\partial{M_{1}}} > 0, ~~~~~
	\partial{v_{j}}/\partial{M_{1}} > 0
	\end{gather}
		
	Correspondingly, the upstream conditions $\rho{}_{1}$ and $T_{1}$ are constant, then:	
	\begin{gather}
	\label{eq:ineq1}
	\partial{\rho{}_{2}}/\partial{v_{j}} > 0 \\
	\label{eq:ineq2}
	\partial{T_{2}}/\partial{v_{j}} > 0
	\end{gather}
		
	These equations show that the temperature $T_{2}$ and density $\rho_{2}$ increase with an increase in the velocity jump. Now, under the assumption of an ideal gas, the total enthalpy deposit per unit area for the shocks can be calculated as:
	\begin{gather}
	\label{eq:enthalpy}
	H_{dep} =  \int\limits_{\underset{v_{z}>0}{t=0s}}^{t=60s}(\dfrac{\gamma{}+1}{\gamma{}-1}p_{2}+\rho{}_{2}\dfrac{v_{z}^{2}}{2})\times{}v_{z}\times{}dt
	\end{gather}
		
	Here $v_{z}$ is the velocity of plasma in the simulations, $R$ is the universal gas constant, $\mu{}$ is the effective molar mass, and $\gamma{}$ is the adiabatic exponent. It is interesting to note that, for the considered shock wave events in the numerical model, $v_{z}$ is correlated with $v_{j}$ ($\partial{}v_{z}/\partial{}v_{j} > 0$), and $v_{j}$ is correlated with $v_{D}$ ($\partial{}v_{j}/\partial{}v_{D} > 0$, see Figures~\ref{figure6}\textit{b} and~\ref{figure6}\textit{c}). Correspondingly, one can deduce that $\partial{}v_{z}/\partial{}v_{D} > 0$. By combining this statement with inequalities in Eq.\ref{eq:ineq1}~and~\ref{eq:ineq2}, we find:
	\begin{gather}
	\partial_{}H_{dep}/\partial{}v_{D} > 0
	\end{gather}
	
	This qualitatively explains why the enthalpy inflow depends on the Doppler shift jump in Figure~\ref{figure5}\textit{d}.
	
	The primary goal of this paper is to investigate the relations between properties of UV spectral emission originating in the upper chromosphere, transition region, and corona, and the shock properties at the same heights. However, it is clear that the perturbations causing the shocks originate deeper in the solar atmosphere. As an example, Figure~\ref{figure7} illustrates propagation of the shock \#17 from the photosphere to the corona. We selected a 2.5\,Mm x 2.5\,Mm square box around the localized enhancement in the synthetic SDO/AIA 335\,{\AA} images corresponding to this shock (red square in Figure~\ref{figure7}\textit{a}). To locate the photospheric origin of the shock, we constructed the time-height diagrams of the horizontally-averaged velocity divergence for various sub-regions within this box. Figure~\ref{figure7}\textit{b} illustrates this diagram for the sub-region marked by the red square in Figures~\ref{figure7}\textit{c-f}. As one can see, there is a clear black ridge in the diagram traceable to the photospheric level. If the time-height diagram is constructed for the region outside the red square box, the ridge cannot be traced back to the photosphere. Therefore, the shock initiation process is closely related to the activity and evolution within this box. As evident in Figures~\ref{figure7}\textit{c-f}, the box contains the small granule and the corresponding intergranular lanes. This granule experiences fast evolution and interaction with other small granules. In particular, the granule area decreases more than twice in a matter of 2 minutes, which is faster than the average lifetime of granules of 8.6\,minutes \citep{Bahng61}. The surrounding intergranular lanes also become broader.
	
	Based on the described dynamics of the granule, we make a conclusion that the considered shock \#17 is initiated by a granule collapse mechanism of excitation of acoustic waves \citep{Skartlien00a}. However, our preliminary investigation indicates that another proposed acoustic wave excitation mechanism~--- the interaction of vortices formed primarily in the intergranular lanes \citep{Kitiashvili11a,Kitiashvili12b}~--- also initiates shocks in the considered simulation series. The conclusion about how often each particular mechanism is predominant to initiate shocks in the solar atmosphere requires a detailed statistical investigation, which is out of scope for this work. Nonetheless, the data analysis procedure applied to the shock \#17 can be repeated for other shocks to localize their photospheric origin and excitation mechanism.
	
	When the shock propagates upward to the corona, it steepens and accelerates. The acceleration is sharp in the transition region, because of the strongest gradients of temperature and, correspondingly, the sound speed. In order to qualitatively understand how the shock passes through the transition region, we interpret our numerical solution in terms of a classical problem of interaction of a shock with a contact discontinuity \citep[e.g.][]{Rozhdestvenskii83a}. Because the shock propagates from a higher density region to a lower density region, the solution represents a composition of an upward-traveling shock, an upward-moving discontinuity, and a downward-traveling rarefaction wave. In our simulations, we clearly observe the shocks passing the transition region and propagating upward. We also observe the local upward motion of the transition region after the shock passage, which is analogous to the upward-moving contact discontinuity. The rarefaction waves are not so evident in the simulations but may correspond to the features moving downward from the point where the shock crosses the transition region in the time-distance diagrams (like in Figure~\ref{figure7}\textit{b}).
	
	There are several physical effects that are not yet included in our analysis. First, the simulations did not include the radiative cooling and heating caused by chromospheric spectral lines that may be important for the chromospheric energy balance \citep{Carlsson12a}. This may be one of the probable reasons why the simulated chromosphere is hotter and denser in comparison with the initial VAL model, and the line intensities are stronger. Second, the simulations do not include effects of the non-equilibrium excitation and ionization. It was shown previously that the relaxation times of Hydrogen and Calcium level populations in the upper chromosphere are of several tens of seconds \citep{Carlsson02a,Wedemeyer-Bohm11a}, which is comparable to the timescales of the shock propagation. Both effects may potentially affect the quantitative estimates in Tables~\ref{table1}\,and\,\ref{table2}.

\section{Conclusions}
\label{Section:conclusions}
	
\begin{figure}[t!]
	\centering
	\includegraphics[width=0.80\linewidth]{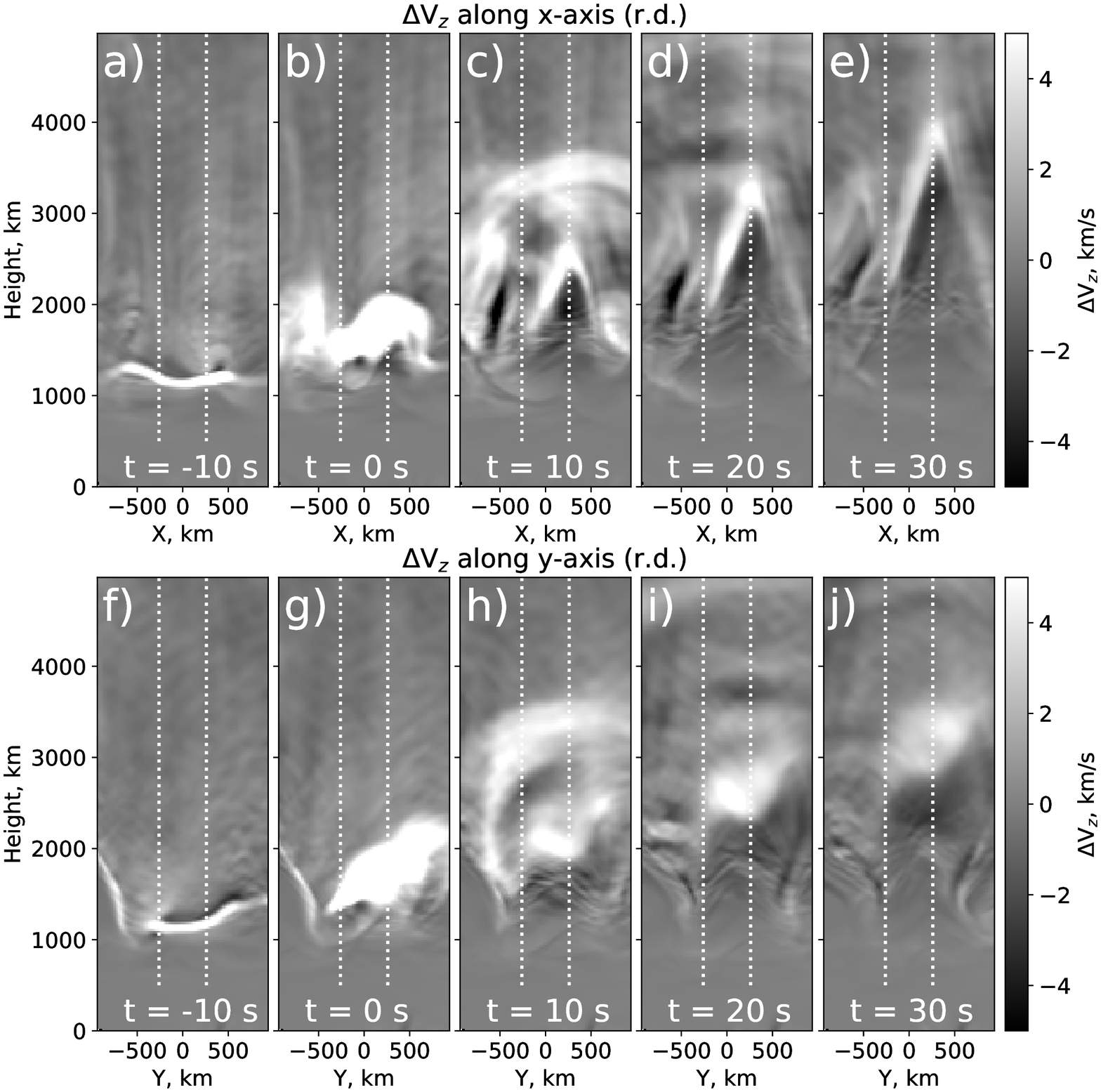} \\
	\includegraphics[width=0.80\linewidth]{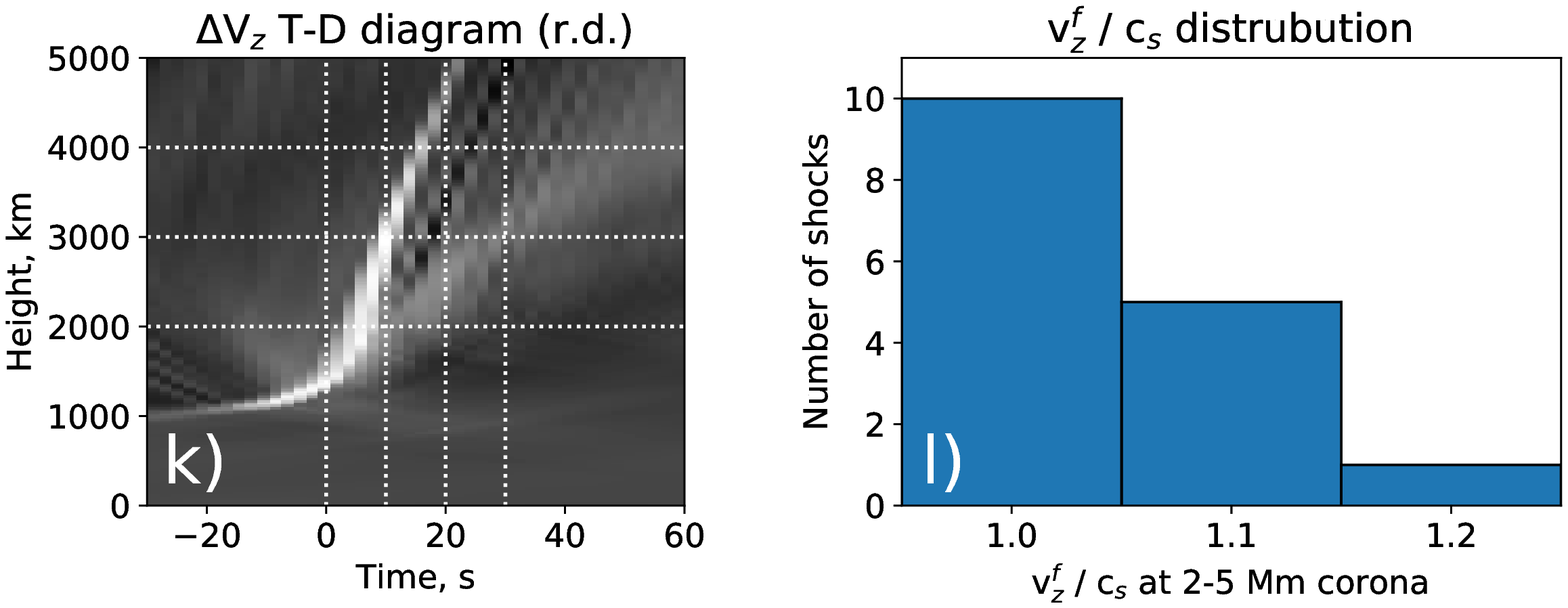}
	\caption{(a-j) evolution of the running difference of vertical velocity ($\Delta{}$v$_{z}$) for a shock event. Panels (a-e) and (f-j) correspond to the slices of the domain by vertical planes along the horizontal X and Y axes respectively. Time moments are marked with respect to the shock reference time. (k) Time-distance diagram of the $\Delta{}$v$_{z}$ integrated horizontally within the white vertical dotted lines marked in panels (a-j). (l) Histogram of the ratios of the shock wave front vertical velocities, $v_{z}^{f}$, determined from time-distance diagrams for the $\Delta{}$v$_{z}$ at heights of 2-5\,Mm, to the average speeds of sound at 2-5\,Mm, $c_{s}$.}
	\label{figure4}
\end{figure}

\begin{figure}[t!]
	\centering
	\includegraphics[width=0.85\linewidth]{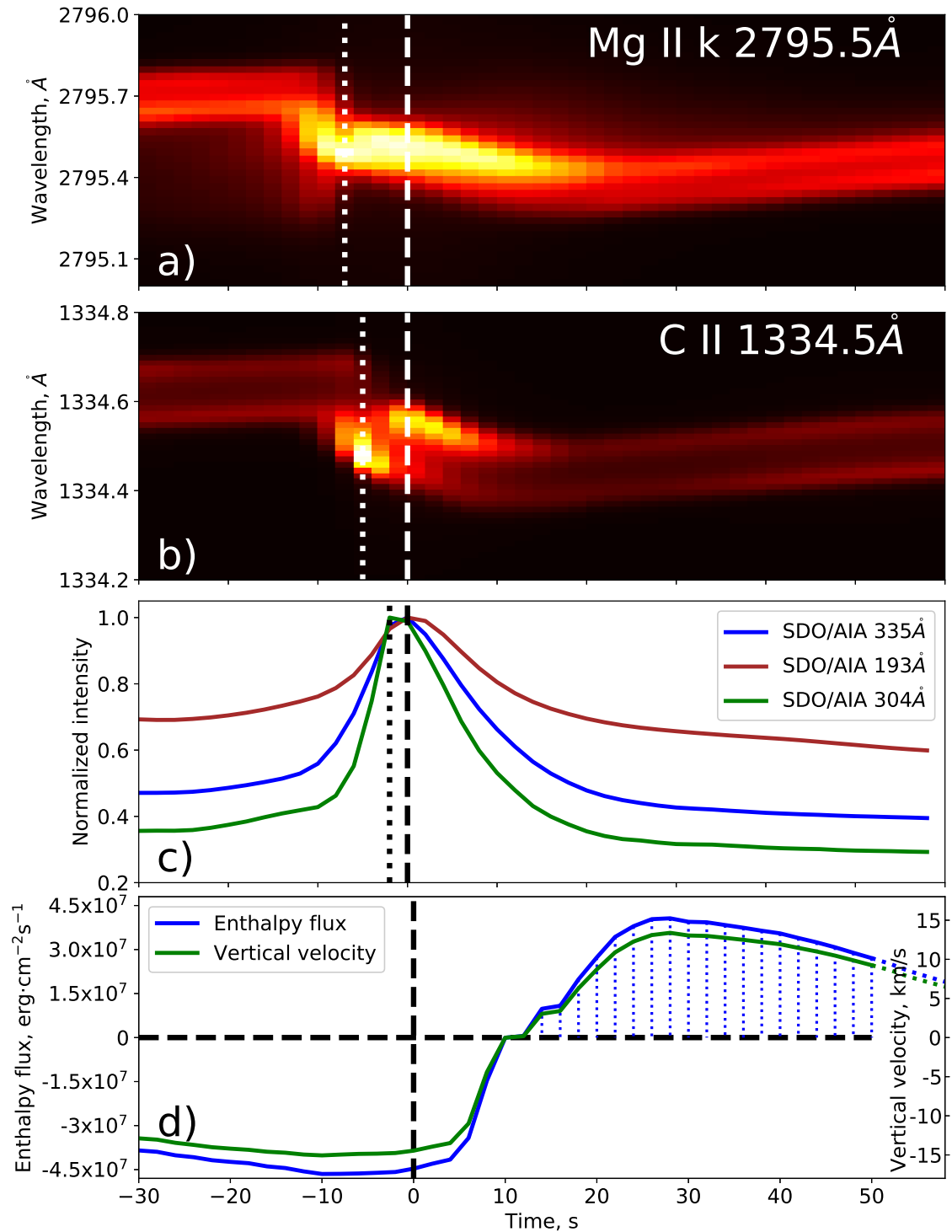}
	\vskip -35pt
	\caption{Evolution of (a) the Mg\,II\,k line spectra, (b) C\,II\,1334.5\,{\AA} line spectra, (c) normalized SDO/AIA emission, and (d) physical parameters in the transition region (at the height corresponding to 5x10$^5$\,K) at the center of the shock event illustrated in Figures~\ref{figure2}-\ref{figure4}. The vertical dashed lines mark the time moment corresponding to the strongest SDO/AIA 335\,{\AA} emission during the shock. The vertical dotted lines mark the time moments of the Mg\,II\,k intensity peak (panel a), C\,II\,1334.5\,{\AA} intensity peak (panel b), and SDO/AIA 304\,{\AA} peak (panel c). The dotted area in panel (d) indicates the positive enthalpy flux.}
	\label{figure3}
\end{figure}

\begin{figure}[t!]
	\centering
	\includegraphics[width=1.0\linewidth]{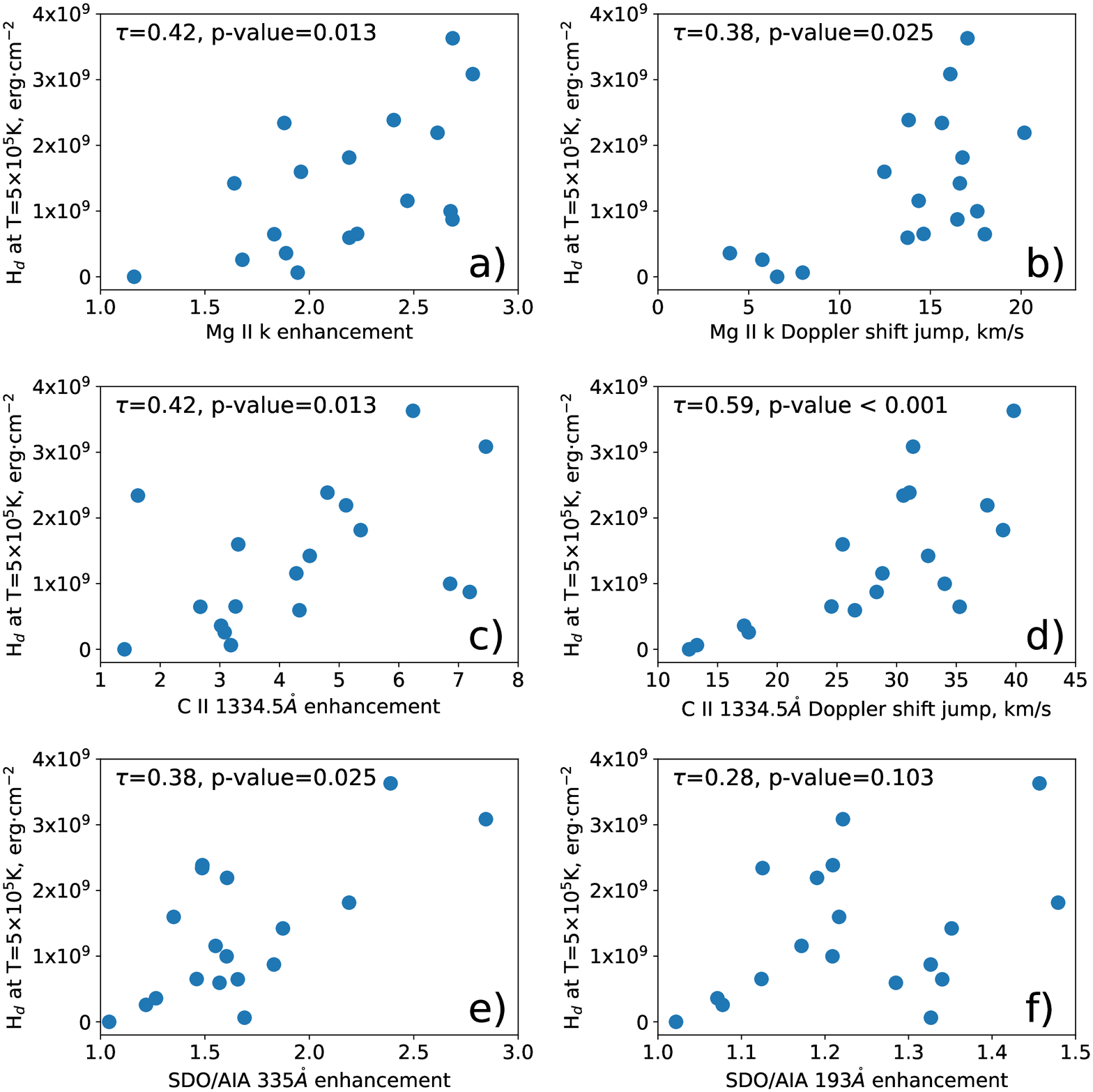}
	\caption{Scatter plots of the enthalpy deposit and (a) the Mg\,II\,k peak intensity enhancement, (b) Mg\,II\,k Doppler shift jump, (c) the C\,II\,1334.5\,{\AA} peak intensity enhancement, (d) C\,II\,1334.5\,{\AA} Doppler shift jump, (e) SDO/AIA 335\,{\AA} emission enhancement, (f) SDO/AIA 193\,{\AA} emission enhancement, derived for the centers of the detected shocks. Corresponding Kendall's $\tau$ coefficients and p-values are indicated at the panels.}
	\label{figure5}
\end{figure}

\begin{figure}[t!]
	\centering
	\includegraphics[width=1.0\linewidth]{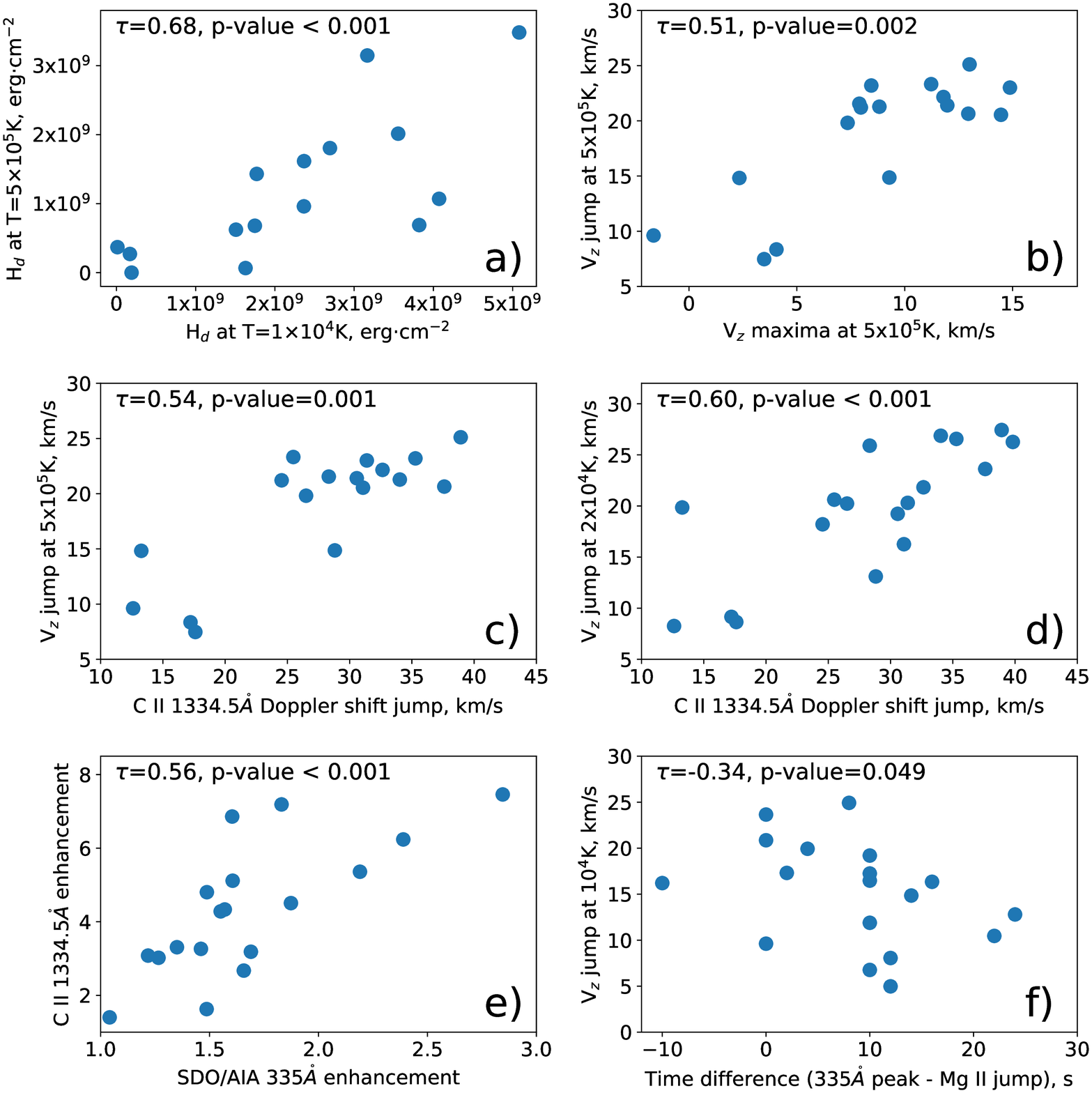}
	\caption{The scatter plots of the (a) enthalpy deposits as the 5x10$^5$\,K and 2x10$^4$\,K heights of the transition region, (b) velocity maximum and jump at the 5x10$^5$\,K height, (c) C\,II\,1334.5\,{\AA} Doppler shift jump at the 5x10$^5$\,K height, (d) C\,II\,1334.5\,{\AA} Doppler shift jump vs the plasma velocity jump at the 2x10$^4$\,K height, (e) the C\,II\,1334.5\,{\AA} peak intensity enhancement vs the SDO/AIA 335\,{\AA} emission enhancement, and (f) time difference between the SDO/AIA 335\,{\AA} strongest emission and the center of the Mg\,II\,1334.5\,{\AA} Doppler shift jump vs the plasma velocity jump at the 2x10$^4$\,K height. Corresponding Kendall's $\tau$ coefficients and p-values are indicated at the panels.}
	\label{figure6}
\end{figure}

\begin{sidewaysfigure}[t!]
	\centering
	\includegraphics[width=1.00\linewidth]{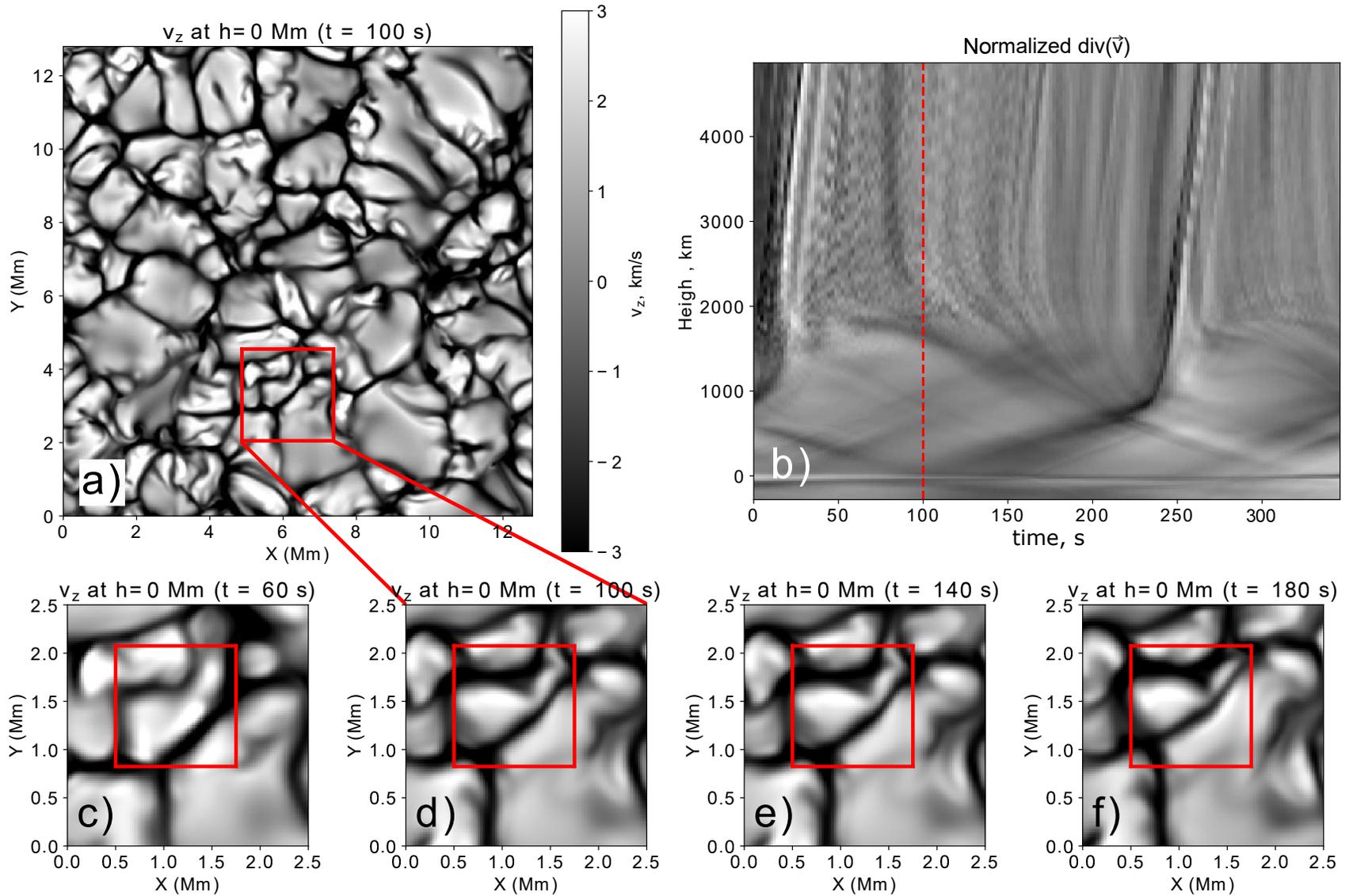}
	\caption{Photospheric footprints of the shock wave \#17. Illustration of the vertical velocity map at  z = 0\,km for t = 100\,s is presented in panel (a). The red square marks the map portion illustrated in panels (c-f) for t = 60\,s, t = 100\,s, t = 140\,s, and t = 180\,s. A time-height diagram of the $div$\,$\vec{v}$ integrated horizontally within the red square marked in panels (c-f) is presented in panel (b). The red dashed line indicates the time moment when the perturbation is initiated at z=0\,km height.}
	\label{figure7}
\end{sidewaysfigure}

	In this study, we analyzed the evolution and properties of shock waves in the solar transition region and corona and their emission using 3D radiative MHD simulations of the quiet Sun performed with the StellarBox RMHD code. The study leads to the following conclusions:
	\begin{itemize}
		\item Shock waves manifest themselves as sharp enhancements of the EUV emission (as observed by SDO/AIA) and the UV C\,II and Mg\,II spectral lines (as observed by IRIS), and are also reflected in the Doppler velocity jumps of these lines;
		\item The Doppler velocity jump of the C II 1334.5\,{\AA} IRIS line and relative enhancement of SDO/AIA 335\,{\AA} emission are among the best proxies for the enthalpy deposited by shocks in the corona with Kendall's $\tau$ correlation coefficients of 0.59 for C\,II line and 0.38 for 335\,{\AA} EUV emission respectively;
		\item The emission of UV lines and EUV bandwidths (e.g., C\,II and SDO/AIA 335\,{\AA} emissions) are correlated with each other during shock wave propagation, which is important for potential observational diagnostics;
		\item All studied shock waves are mostly hydrodynamic and have ratios of shock front vertical speeds to the average speed of sound, $v_{z}^{f}$/$c_{s}$, in the range of 1.0-1.2 at heights of 2-5\,Mm. The shock waves also result in disturbances of physical parameters at a height of 4\,Mm, which are observed as spherically-shaped perturbations (Fig.~\ref{figure4});
		\item The empirical correlation of the C\,II\,1334.5\,{\AA} Doppler shift jumps and deposited enthalpy is in qualitative agreement with the Rankine-Hugoniot relations.
	\end{itemize}
	
	We conclude that the current study reveals the possibility of analyzing the enthalpy transported by shock waves into the solar corona by utilizing remote sensing observations. Shocks have signatures in the spectra of UV lines observed by IRIS and formed in the upper chromosphere and transition region, as well as in EUV emission of the hot solar corona observed by SDO/AIA. This demonstrates the possibility of studying the shocks and analyzing their properties with currently operational satellites.

\acknowledgments

We acknowledge the NASA Ames Research Center for the possibility to use the computational resources and support with the installation and performance optimization of the utilized codes. The research was supported by NASA grants HSR NNX12AD05A, NNX14AB68G, NNX16AP05H, NSF grant 1916509.

\bibliographystyle{aasjournal}

\bibliography{StellarBox_shockwaves}

\newpage

\newpage

\newpage

\newpage

\newpage

\newpage
\begin{sidewaystable}
	\centering
	\caption{{Physical characteristics and synthetic emission parameters of shocks (part 1).}}
	\label{table1}
	\scriptsize
	\begin{tabular}{MMMMMMMM}
		\hline
		Event number	&	Enthalpy deposit at T=5$\times$10$^{5}$\,K, 10$^{8}$~erg$\cdot$cm$^{-2}$	&	Mg II k enhancement	&	Mg II k Doppler jump, km/s	&	C II 1334.5$\AA$ enhancement	&	C II 1334.5$\AA$ Doppler jump, km/s	&	SDO/AIA 335$\AA$ enhancement	&	SDO/AIA 193$\AA$ enhancement	\\
		\hline
		1	&	16.0	&	1.96	&	12.5	&	3.31	&	25.5	&	1.35	&	1.22	\\
		2	&	18.1	&	2.19	&	16.8	&	5.36	&	38.9	&	2.19	&	1.48	\\
		3	&	5.94	&	2.19	&	13.7	&	4.33	&	26.5	&	1.57	&	1.28	\\
		4	&	2.59	&	1.68	&	5.75	&	3.08	&	17.6	&	1.22	&	1.08	\\
		5	&	11.6	&	2.47	&	14.4	&	4.28	&	28.8	&	1.55	&	1.17	\\
		6	&	0.00	&	1.16	&	6.56	&	1.40	&	12.6	&	1.04	&	1.02	\\
		7	&	6.52	&	2.23	&	14.6	&	3.26	&	24.5	&	1.46	&	1.12	\\
		8	&	23.4	&	1.88	&	15.6	&	1.63	&	30.6	&	1.49	&	1.13	\\
		9	&	3.59	&	1.89	&	3.96	&	3.02	&	17.2	&	1.27	&	1.07	\\
		10	&	21.9	&	2.61	&	20.2	&	5.11	&	37.6	&	1.61	&	1.19	\\
		11	&	9.98	&	2.68	&	17.6	&	6.86	&	34.0	&	1.60	&	1.21	\\
		12	&	30.9	&	2.78	&	16.1	&	7.46	&	31.4	&	2.85	&	1.22	\\
		13	&	8.73	&	2.69	&	16.5	&	7.19	&	28.3  	&	1.83	&	1.33	\\
		14	&	6.48	&	1.83	&	18.0	&	2.67	&	35.3	&	1.66	&	1.34	\\
		15	&	23.9	&	2.40	&	13.8	&	4.80	&	31.1 	&	1.49	&	1.21	\\
		16	&	0.63	&	1.94	&	7.97	&	3.19	&	13.3 	&	1.69	&	1.33	\\
		17	&	14.2	&	1.64	&	16.6	&	4.51	&	32.6 	&	1.87	&	1.35	\\
		18	&	36.3	&	2.69	&	17.0	&	6.24	&	39.8	&	2.39	&	1.46	\\
		\hline
	\end{tabular}
\end{sidewaystable}

\newpage

\newpage
\begin{sidewaystable}
	\centering
	\caption{Physical characteristics and synthetic emission parameters of shocks (part 2).}
	\label{table2}
	\scriptsize
	\begin{tabular}{LLLLLLL}
		\hline
		Event number	&	$^{*}$Enthalpy deposit at T=5$\times$10$^{5}$\,K, 10$^{8}$~erg$\cdot$cm$^{-2}$	&	Enthalpy deposit at T=1$\times$10$^{4}$\,K, 10$^{8}$~erg$\cdot$cm$^{-2}$	&	Vertical velocity jump at T=5$\times$10$^{5}$\,K, km/s	&	Vertical velocity jump at T=2$\times$10$^{4}$\,K, km/s	&	Vertical velocity maxima at T=5$\times$10$^{5}$\,K, km/s	&	Time difference between SDO/AIA 335$\AA$ emission peak and Mg\,II\,k Doppler jump, s	\\
		\hline
		1	&	18.1	&	26.9	&	20.6	&	23.3	&	11.0	&	2		\\
		2	&	20.1	&	35.6	&	27.4	&	25.1	&	12.4	&	8		\\
		3	&	6.23	&	15.1	&	20.2	&	19.8	&	6.48	&	-10	\\
		4	&	2.70	&	1.70	&	8.65	&	7.47	&	2.41	&	10		\\
		5	&	14.3	&	17.7	&	13.1	&	14.9	&	5.41	&	0		\\
		6	&	0.0	    &	1.91	&	8.26	&	9.62	&	-0.54	&	12		\\
		7	&	6.80	&	17.5	&	18.2	&	21.2	&	6.15	&	14		\\
		8	&	34.8	&	50.8	&	19.2	&	21.4	&	10.1	&	22		\\
		9	&	3.69	&	0.13	&	9.15	&	8.35	&	2.33	&	12		\\
		10	&	29.5	&	61.7	&	23.6	&	20.6	&	11.0	&	10		\\
		11	&	10.7	&	40.7	&	26.9	&	21.3	&	8.67	&	10		\\
		12	&	42.2	&	44.2	&	20.3	&	23.0	&	13.1	&	16		\\
		13	&	9.61	&	23.7	&	25.9	&	21.6	&	7.70	&	4		\\
		14	&	6.89	&	38.2	&	26.6	&	23.2	&	7.74	&	0		\\
		15	&	31.5	&	31.7	&	16.3	&	20.5	&	12.9	&	10		\\
		16	&	0.66	&	16.3	&	19.9	&	14.8	&	3.85	&	24		\\
		17	&	16.2	&	23.7	&	21.8	&	22.2	&	9.63	&	10		\\
		18	&	43.1	&	56.4	&	26.3	&	30.7	&	15.0	&	0		\\
		\hline
	\end{tabular} \\
	\vspace{2ex}
	{\raggedright $^{*}$ The potential energy with respect to T=1$\times$10$^{4}$\,K height is added\par}
\end{sidewaystable}

\newpage

\newpage

\end{document}